\documentclass{iaus238}
\usepackage{graphicx}

\title[Evidence for stellar-mass black holes] 
{Observational evidence for stellar-mass \\[\affilskip] black holes} 

\author[J. Casares]   
{Jorge Casares}

\affiliation{$^1$~Instituto de Astrof\'\i{}sica de Canarias, 38200 -- La Laguna, Tenerife, Spain\\[\affilskip]
email: jcv@iac.es}

\pubyear{2006}
\volume{238}  
\pagerange{001--999}
\date{??? and in revised form ???}
\jname{Black Holes: from Stars to Galaxies -- across the Range of Masses}
\editors{V. Karas \& G. Matt, eds.}
\begin{document}

\maketitle

\begin{abstract}
Radial velocity studies of X-ray binaries provide the most solid evidence 
for the existence of stellar-mass black holes. We currently have 20 confirmed 
cases, with dynamical masses in excess of 3 M$_{\odot}$. Accurate masses have 
been obtained for a subset of systems which gives us a hint at the mass 
spectrum of the black hole population. This review summarizes the history of 
black hole discoveries and presents the latest results in the field. 
\end{abstract}

\firstsection 
\section{Introduction}
Theoretical black holes (BHs) have been considered for almost a century, 
although their first observational evidence was not found until quite recently, 
during the past 3 decades. Since then BHs have become essential ingredients 
in the construction of modern astrophysics on all scales, from binaries to 
galaxies and AGNs. It is stellar-mass BHs that offer us the best opportunity 
to study these objects in detail. Their implications are wide ranging, from 
late evolution of massive stars, supernovae models, production of high-energy 
radiation, relativistic outflows, chemical enrichment in the Galaxy, etc. 
But still the most compelling evidence for the existence of stellar-mass BHs
relies on dynamical arguments, which will be the focus of this review. Section 
2 summarizes the first observations of historical BH ``candidates'', whereas 
Sect. 3 reviews the most recent discoveries. In Sect. 4 we deal with BH 
demography and their mass distribution. Finally, we present our 
conclusions in Section 5.   

\section{Early discoveries: the first BH ``candidates''}

In the late 1960s X-ray detectors onboard satellites revolutionized 
astronomy  with the discovery of an unexpected population of luminous X-ray 
sources in the Galaxy. The energetics (with 
$L_{\rm x}\sim L_{\rm Edd}$) together with their short timescale  
variability (down to milliseconds) lend support 
to an interacting binary model where X-rays are supplied by accretion onto a 
collapsed object (\cite{shk67}). It is now well stablished that there are two 
main populations of X-ray binaries:  
the high mass X-ray binaries (HMXBs), containing O-B supergiant donor stars, 
and the low mass X-ray binaries (LMXBs), with typically short orbital periods 
and K-M donors. The optical flux in LMXBs is triggered by reprocessing 
of the X-rays into the accretion disc whereas in HMXBs it is dominated by the 
hot supergiant star (see reviews in Charles \& Coe 2006 and \cite{mcc06}). 

Therefore, it was not surprising that one of the first optical counterparts to 
be identified was the 9th magnitude supergiant star 
HD 226868, associated with the HMXB Cyg X-1. But, remarkably,  	
it showed radial velocity variations which made it a prime candidate for 
a stellar-mass BH (\cite{web72}, \cite{bolton72}). The supergiant star was 
shown to move with a velocity amplitude of $\sim$64 km s$^{-1}$ (later refined 
to 75 km s$^{-1}$) in a 5.6 day orbit due to the gravitational influence of an 
unseen companion (see Fig. 1). 
The orbital period $P_{\rm orb}$ and the radial velocity amplitude $K$, 
combine in the mass function equation 
$f(M_{\rm x}) =  K^{3} P_{\rm orb}/2 \pi G = M^{3}_{\rm x}\sin^{3} 
i/\left(M_{\rm x} + M_{\rm c}\right)^2$ 
which relates the mass of the compact object $M_{\rm x}$ with that of the 
companion star $M_{\rm c}$ and the inclination angle $i$. The mass function 
$f(M_{\rm x})$ is a lower limit to 
$M_{\rm x}$ and, for the case of Cyg X-1 is 0.25 M$_{\odot}$. The key 
factor here is $M_{\rm c}$ which, for a HMXB, is a large number and has a wide 
range of uncertainty. If the optical star were a normal O9.7Iab its 
mass would be $\sim$ 33 M$_{\odot}$ which, for an edge-on orbit
($i=90^{\circ}$), would imply a compact object of $\sim$ 7 M$_{\odot}$. 
However, the optical star is likely to be undermassive for its spectral type 
as a result of mass transfer and binary evolution, as has been shown to be the
case in several neutron star binaries (e.g. \cite{rap83}).  
In fact, it could be undermassive by as much as a factor of 3 given the 
uncertainty in distance, $\log g$ and $T_{\rm eff}$. 
A plausible lower limit of 10 M$_{\odot}$, combined with an upper limit to the 
inclination of 60$^{\circ}$, based on the absence of X-ray eclipses, leads to 
a compact object of  $>$4 M$_{\odot}$ (\cite{bolton75}).

\begin{figure}[ht]
\includegraphics[width=0.6\textwidth]{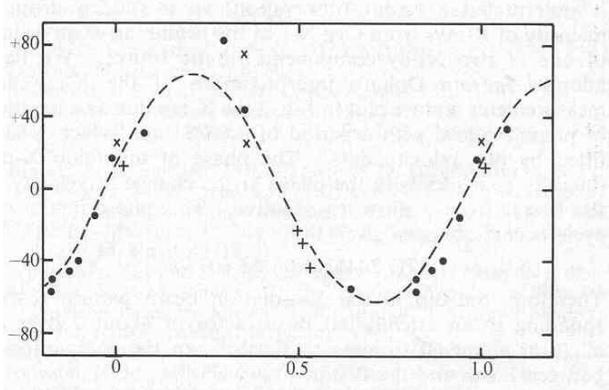}
\hfill
\parbox[b]{0.39\textwidth}{\caption{Radial velocity curve of 
HD 226868, the O9.7Iab companion star in the HMXB Cyg X-1, 
folded on the 5.6 day orbital period. After Webster \& 
Murdin (1972).\newline}} 
\label{fig:fig1}
\end{figure}

The importance of this result rests on  the fact that there is a 
maximum mass for neutron stars (NS) to be stable against gravitational 
collapse (\cite{oppen39}). This maximum depends on the 
equation of state (EoS), which is uncertain in the high density regime because
of the poorly constrained many-body interactions.   
However, \cite{rho74} showed that an upper limit of $\sim$3.2 M$_{\odot}$ can 
be derived assuming that causality holds beyond densities where the EoS starts 
to be uncertain which, at the time, was 1.7 times  the density of nuclear matter 
$\rho_{nm}=2.7\times10^{14}$ g cm$^{-3}$.  
More recently, a new limit of 2.9 M$_{\odot}$ was obtained using modern EoS 
which are accurate up to $2 \times \rho_{nm}$ (\cite{kalo96}). This can 
be further boosted by up to $\sim$25\% if the NS rotates close to break-up
(\cite{fried87}).  
Therefore, the compact object in Cyg X-1, with 
$M_{\rm x} \ge$ 4 M$_{\odot}$, is a very strong BH candidate.  

In 1975 the satellite Ariel V detected A0620-00, a new X-ray source which
displayed an increase in X-ray flux from non-detection to a record of 
$\sim$50 Crab. It belongs to the class of X-ray transients (XRTs, also 
called X-ray Novae), a subclass of LMXBs which undergo dramatic episodes of 
enhanced mass-transfer or ``outbursts" triggered by viscous-thermal 
instabilities in the disc (e.g. \cite{king99}). During outburst, the 
companion remains undetected because it is totally 
overwhelmed by the intense optical light from the X-ray heated disc.  
However, the X-rays switch off after a few months of activity, the 
reprocessed flux drops several magnitudes into quiescence and the companion 
star becomes the dominant source of optical light. 
This offers a very special opportunity to perform radial velocity studies of 
the cool companion and unveil the nature of the compact star. 
The first detection of the companion in A0620-00 revealed a mid-K star 
moving in a 7.8 hr period with velocity amplitude of 457 km s$^{-1}$. The 
implied mass function was 3.2 $\pm$ 0.2 M$_{\odot}$, the largest ever 
measured (\cite{mcc86}). An absolute 3$\sigma$ lower limit to the mass of the 
compact star of 3.2 M$_{\odot}$ was established by assuming a very conservative 
low-mass companion of 0.25 M$_{\odot}$ and $i<85^{\circ}$, based on the lack of 
X-ray eclipses. This exceeds the maximum mass allowed for a stable NS and hence 
it also became a very compelling case for a BH.  

\section{From ``candidates'' to confirmed BHs}

In the 1980s there was a hot debate about the real existence of BHs. 
On one hand there were 3 strong candidates, the HMXBs Cyg X-1 and LMC X-3 and 
the transient  LMXB A0620-00, all with lower limits to M$_{\rm x}$ very close 
to the maximum mass for NS stability. 
On the other, alternative scenarios were proposed to avoid the need for BHs 
such as multiple star systems (\cite{bah74}) or non-standard models invoking 
exotic EoS for condensed matter. An example of the latter are Q-stars, where 
neutrons and protons are confined by the strong force rather than gravity and 
can be stable 
for larger masses (\cite{bah90}). In this context, it was proposed that the 
{\it ``holy grail in the search for BHs is a system with a mass function that 
is plainly 5 M$_{\odot}$ or greater''} (\cite{mcc86b}). 

\begin{figure}[ht]
\includegraphics[width=0.45\textwidth,angle=-90]{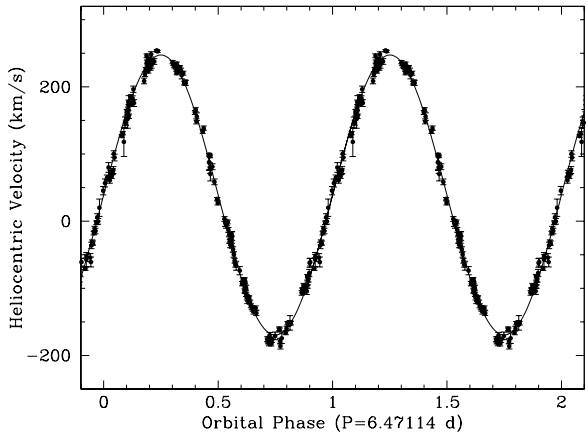}
\hfill\par\vspace*{-9em}\hfill
\parbox[b]{0.35\textwidth}{\caption{Radial velocity curve of the K0 companion in the transient LMXB 
V404 Cyg during quiescence. This graph contains velocity points obtained
between 1991 and 2005.}}
\label{fig:fig2}
\vspace*{3em}
\end{figure}

In 1989, the X-ray satellite Ginga discovered a new XRT in 
outburst named GS 2023+338 (=V404 Cyg). Its X-ray properties drew considerable 
attention because of the exhibition of a possible luminosity saturation at 
L$_{\rm x}\simeq10^{39}$ erg s$^{-1}$ and dramatic variability 
(\cite{zycki99}). 
Spectroscopic analysis during quiescence revealed a K0 star moving with a
velocity amplitude of 211 km s$^{-1}$ in a 6.5 day orbit (see Fig. 2).  
The mass function implied by these numbers is 6.3 $\pm$ 0.3 M$_{\odot}$, 
and hence the compact object must be more massive than 6 M$_{\odot}$, 
independent of any assumption on $M_{\rm c}$ and $i$ (\cite{casa92}). This 
remarkable result established V404 Cyg as the ``holy grail'' BH for almost a 
decade. Since then, many other BHs have been unveiled through dynamical studies 
of XRTs in quiescence, seven others with mass functions also in excess of 5 
M$_{\odot}$. This has been possible thanks to the improvement in spectrograph
performance on a new generation of 10-m class telescopes over the last decade. 

\begin{table}\def~{\hphantom{0}}
  \begin{center}
  \caption{Confirmed black holes and mass determinations\vspace*{3mm}}
  \label{tab:bh}
  \begin{tabular}{lccccc}\hline
 System &  $P_{\rm{}orb}$ &  $f(M)$ & Donor  &  Classification & $M_{\rm x}$~$^{\dagger}$ \\
 &   [days] &  [$M_{\odot}$] &  Spect. Type & &  [$M_{\odot}$] \\ \hline
GRS 1915+105$^a$    &     33.5    &    9.5 $\pm$ 3.0        &    K/M III   & LMXB/Transient   &   14 $\pm$ 4   \\
V404 Cyg        &      6.471  &   6.09 $\pm$ 0.04       &    K0 IV     &      ,,          &   12 $\pm$ 2   \\
Cyg X-1         &      5.600  &  0.244 $\pm$ 0.005      &    09.7 Iab  &  HMXB/Persistent &   10 $\pm$ 3   \\
LMC X-1         &      4.229  &   0.14 $\pm$ 0.05       &    07 III    &      ,,          &   $>$ 4         \\
XTE J1819-254   &      2.816  &   3.13 $\pm$ 0.13       &    B9 III    &  IMXB/Transient  &  7.1 $\pm$ 0.3 \\ 
GRO J1655-40    &      2.620  &   2.73 $\pm$ 0.09       &    F3/5 IV   &      ,,          &  6.3 $\pm$ 0.3 \\
BW Cir$^b$ &    2.545  &   5.74 $\pm$ 0.29       &    G5 IV     &  LMXB/Transient  &    $>$ 7.8     \\	 
GX 339-4        &      1.754  &   5.8  $\pm$ 0.5        &     --       &      ,,          &                \\
LMC X-3         &      1.704  &   2.3  $\pm$ 0.3        &    B3 V      &  HMXB/Persistent &  7.6 $\pm$ 1.3 \\
XTE J1550-564   &      1.542  &   6.86 $\pm$ 0.71       &    G8/K8 IV  &  LMXB/Transient  &  9.6 $\pm$ 1.2 \\
4U 1543-475     &      1.125  &   0.25 $\pm$ 0.01       &    A2 V      &  IMXB/Transient  &  9.4 $\pm$ 1.0 \\
H1705-250       &      0.520  &   4.86 $\pm$ 0.13       &    K3/7 V    &  LMXB/Transient  &    6 $\pm$ 2   \\
GS 1124-684     &      0.433  &   3.01 $\pm$ 0.15       &    K3/5 V    &      ,,          &  7.0 $\pm$ 0.6 \\
XTE J1859+226$^c$  &  0.382  &   7.4  $\pm$ 1.1        &     --       &      ,,          &                \\
GS2000+250      &      0.345  &   5.01 $\pm$ 0.12       &    K3/7 V    &      ,,          &  7.5 $\pm$ 0.3 \\
A0620-003       &      0.325  &   2.72 $\pm$ 0.06       &    K4 V      &      ,,          &   11 $\pm$ 2   \\
XTE J1650-500   &      0.321  &   2.73 $\pm$ 0.56       &    K4 V      &      ,,          &                \\
GRS 1009-45     &      0.283  &   3.17 $\pm$ 0.12       &    K7/M0 V   &      ,,          &  5.2 $\pm$ 0.6 \\
GRO J0422+32    &      0.212  &   1.19 $\pm$ 0.02       &    M2 V      &      ,,          &    4 $\pm$ 1   \\
XTE J1118+480   &      0.171  &   6.3  $\pm$ 0.2        &    K5/M0 V   &      ,,          &  6.8 $\pm$ 0.4 \\\hline
  \end{tabular}
  \end{center}
$^{\dagger}$ Masses compiled by \cite{oro03} and \cite{char06}. \\
$^a$ New photometric period of 30.8$\pm$0.2 days recently reported by 
\cite{neil06}. The implied mass function, assuming constant velocity amplitude, 
would be 8.7 M$_{\odot}$. \\ 
$^b$ Updated after \cite{casa07}. \\
$^c$ Period is uncertain, with another possibility at 0.319 days (see
\cite{zur02}). This would drop the mass function to 6.18 M$_{\odot}$.  
\end{table}

Table 1 presents the current list of 20 confirmed BHs based on dynamical 
arguments, ordered by orbital period. The case of GRS 1915+105 is noteworthy, 
not only because of its long orbital period and large mass function; also 
because IR spectroscopy was needed to overcome the $>25$ magnitudes of 
optical extinction and reveal the radial velocity curve of the 
companion star (\cite{grei01}). It should be noted that, although some systems 
have mass functions $<$ 3 M$_{\odot}$, solid constraints on the inclination 
and/or M$_{\rm c}$ can be set  which result in 
M$_{\rm x} >$ 3 M$_{\odot}$. The great majority of BHs are transients and only 
3 show persistent behaviour, the HMXBs Cyg X-1, LMC X-1 and LMC  X-3.  
From the list we also note that a new class of transient X-ray binaries with 
A-F companions is starting to 
emerge, the so-called Intermediate Mass X-ray Binaries (IMXBs). It has 
been proposed that LMXBs may descend from IMXBs through a phase of thermal
mass-transfer (\cite{pfahl03}), although it is still unclear whether IMXBs and 
LMXBs represent an evolutionary sequence. 

\subsection{A novel technique: fluorescence emission from the irradiated donor}

In addition to the dynamical BHs, there are $\sim20$ other BH candidates 
based on their X-ray temporal and spectral behaviour (\cite{mcc06}). 
Unfortunately, they have never been seen in quiescence, or they simply become 
too faint for an optical detection of the companion star. However, a new 
strategy was devised to allow the extraction of dynamical information in 
these systems during their X-ray active states. It 
utilises narrow high-excitation emission lines powered by irradiation on the 
companion star, in particular the strong CIII and fluorescence NIII lines from 
the Bowen blend at $\lambda\lambda$4630-40 (see Fig. 3). 
This technique was first applied to the NS LMXB Sco X-1 and the Doppler 
shift of the CIII/NIII lines enabled the motion of the donor star 
to be traced for the first time (\cite{stee02}). 
This was also attempted during the 2002 outburst of the BH candidate GX 339-4,  
using high-resolution spectroscopy to resolve the sharp NIII/CIII lines. 
The right panel in Figure 3 shows the radial velocity
curve of the NIII lines folded on  the 1.76 day orbital period.   
The orbital solution yields a velocity semi-amplitude of 317 km s$^{-1}$ 
which defines a strict lower limit to the velocity amplitude of the companion 
star because these lines arise from the irradiated 
hemisphere and not the center of mass of the donor. Therefore, a solid lower 
bound to the mass function is 5.8 M$_{\odot}$ which provides compelling 
evidence for a BH in GX 339-4 (\cite{hynes03}). 
And most important, this technique opens an avenue to extract dynamical 
information from new XRTs in outburst and X-ray persistent LMXBs, which 
hopefully will help increase the number of BH discoveries.  

\begin{figure}[ht]
\begin{center}
\begin{picture}(250,190)(50,30)
\put(0,0){\includegraphics{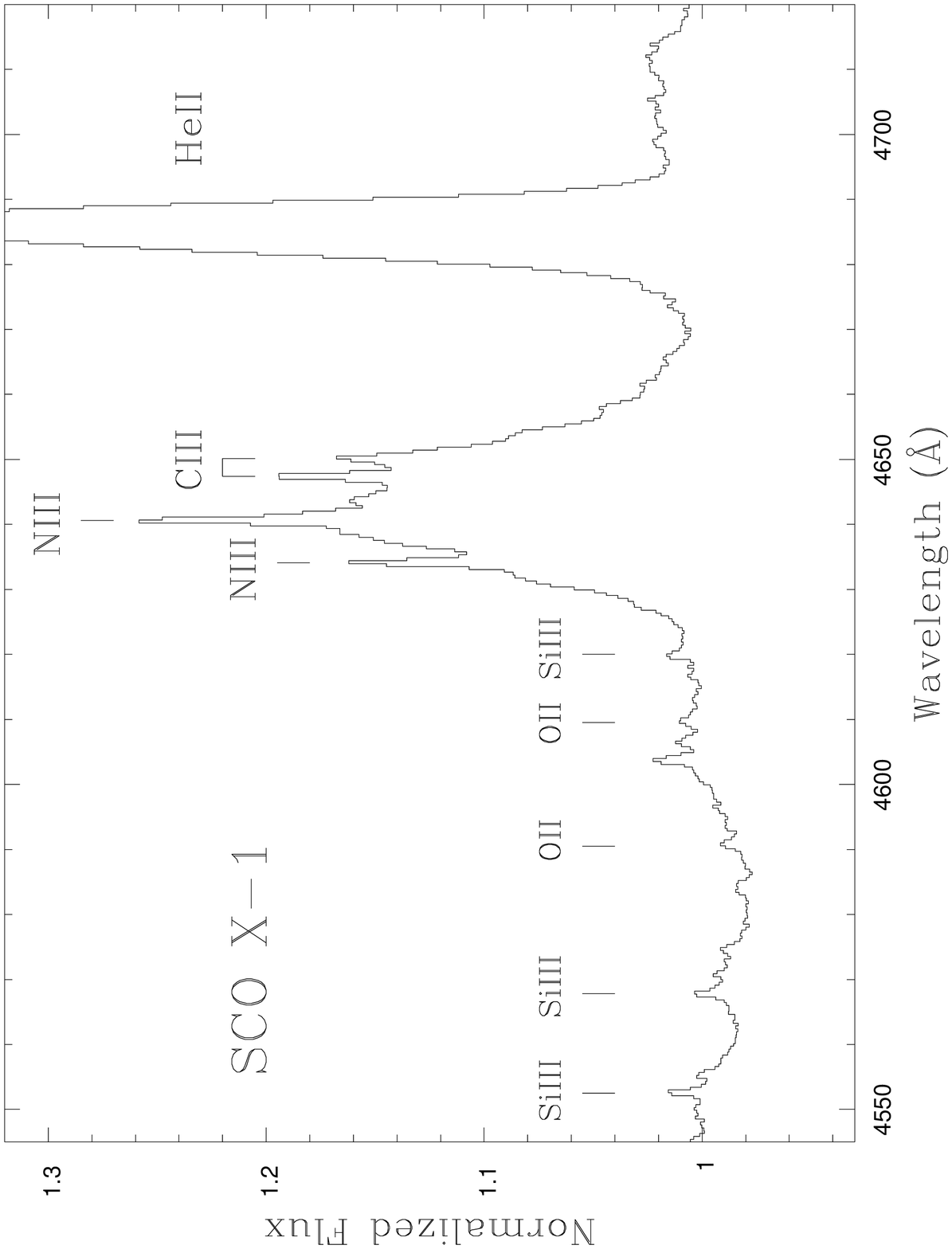}}
\put(0,0){\includegraphics{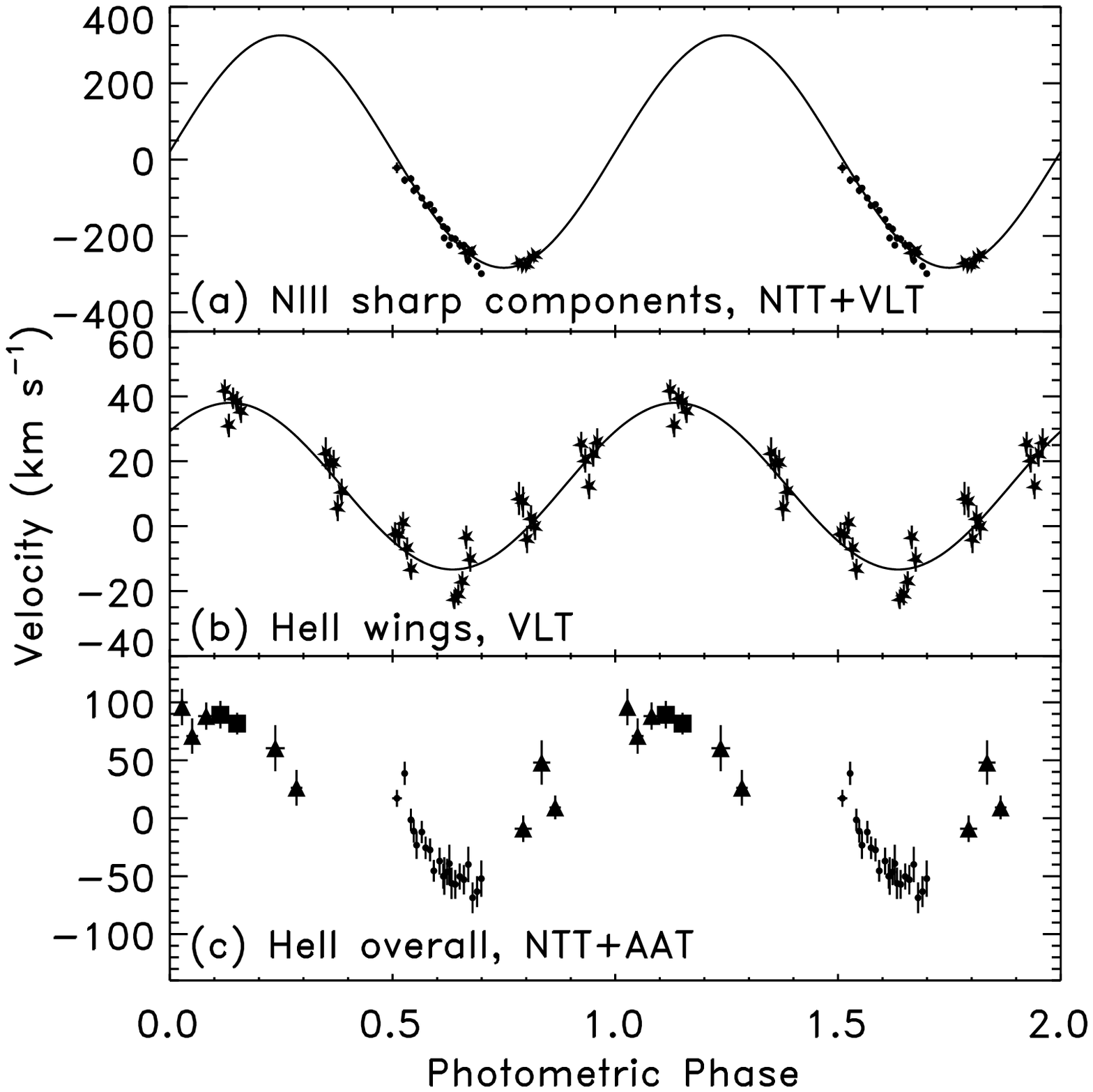}}
\noindent
\end{picture}
\end{center}
{\caption{Detecting the companion star in X-ray active LMXBs. Left: the main 
high excitation emission lines due to by irradiation of 
the donor star in Sco X-1. Adapted from Steeghs \& Casares (2002). 
Right: Radial velocities of the NIII lines (top), 
HeII $\lambda$4686 (bottom) and the wings of HeII $\lambda$4686 (middle) 
in GX 339-4  folded on the 1.76 day orbital period. 
After Hynes {\etal} (2003).}} 
\label{fig:fig3}
\end{figure}

\subsection{Further BH signatures}

Aside from dynamical arguments, there are other observations that 
lend support to the absence of a solid surface and, hence, the BH nature of 
these accreting compact objects, namely: 

\begin{itemize}

\item{} Lack of pulses and Type I X-ray bursts. This has been quantified by 
\cite{rem06} using observations of dynamical BHs over 9 years of RXTE 
data. The probabiliy that the non-detection of bursts were consistent with a 
solid surface is found to be $\sim2\times10^{-7}$. 

\item{} The classic colour-colour diagram of X-ray binaries at high accretion 
rates shows a clear separation in the evolution of NS and BH binaries 
(\cite{done99}). This has been ascribed to the presence of a boundary layer in 
NS which gives rise to an additional thermal component in the spectrum and 
drags NS outside the BH region.

\item{} For a given orbital period, quiescent BH binaries are $\sim$ 100 times 
dimmer than quiescent NS binaries (\cite{menou99a}). This difference is 
interpreted as thermal radiation from the NS surface which, for BHs, is 
advected through the event horizon. 

\end{itemize}

\section{BH demography}

Despite the handful of confirmed BHs one can try to account for 
selection effects and estimate the size of the underlying galactic 
population. To start with, dynamical studies of XRTs indicate that about 
75\% contain BHs, i.e. M$_{\rm x}> 3$ M$_{\odot}$. 
Also, the extrapolation of the number of BH XRTs detected since 1975, with 
outburst duty cycles of $\sim10-100$ years, 
suggests that there is a dormant population of $\sim10^3$ BH binaries
(\cite{romani98} and references therein). 
This is likely to be an underestimate if one accounts for systems with longer
recurrence times and a likely population of faint persistent BH LMXBs 
(\cite{menou99b}).    
Even  so, incidentally, these numbers are  in reasonable agreement with 
recent population synthesis calculations of BH binaries (\cite{yun06}). 
On the other hand, stellar evolution models predict a population 
of about $10^9$ stellar-mass BHs in the Galaxy (\cite{brown94}).
Therefore, our observed sample of BH XRTs is just the tip of the iceberg of a 
large hidden population of which nothing is known. Is it still 
possible to extract some meaningful information from such a limited sample?

\begin{figure}[ht]
\begin{center}
\begin{picture}(250,200)(50,30)
\put(0,0){\includegraphics{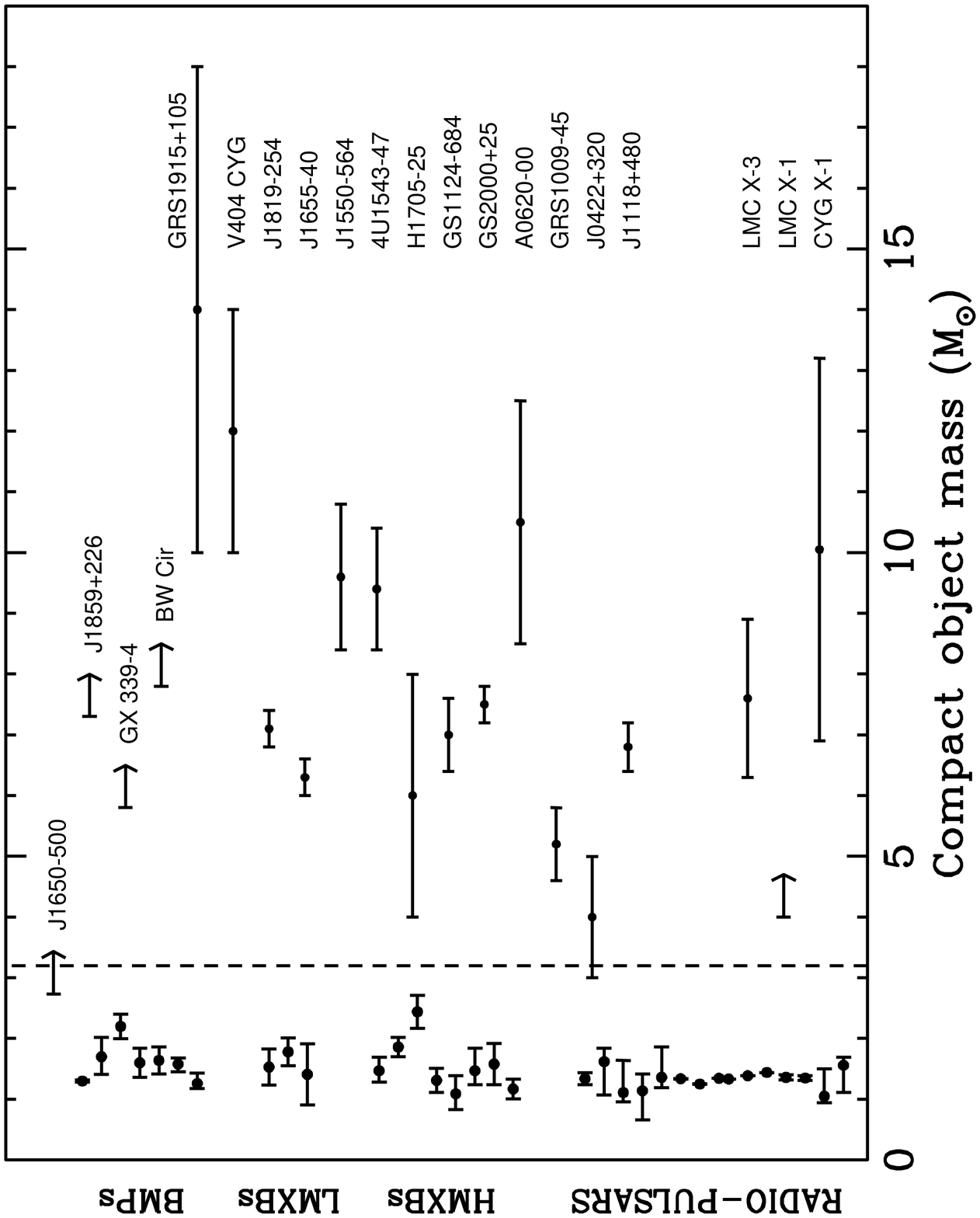}}
\put(0,0){\includegraphics{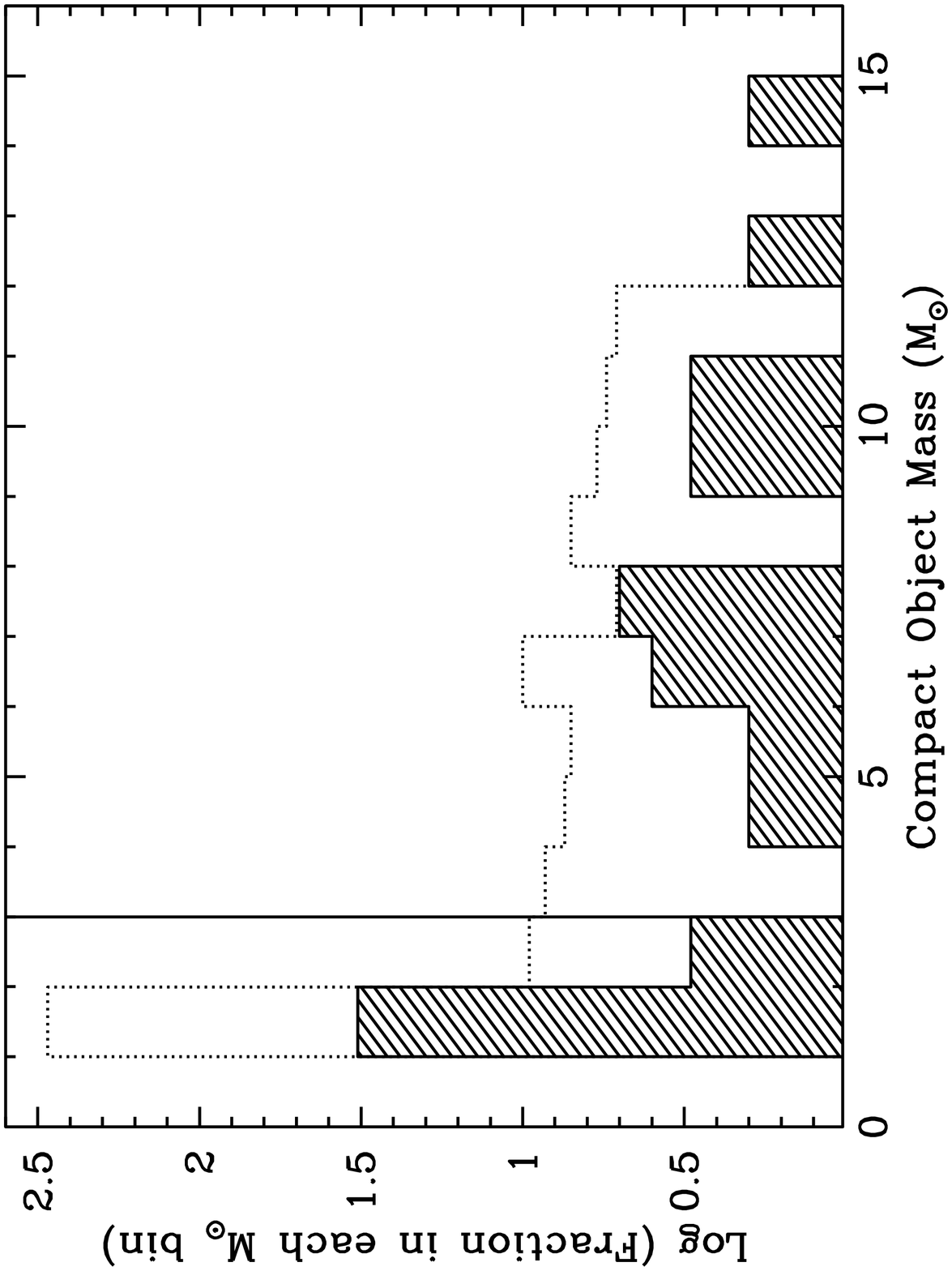}}
\noindent
\end{picture}
\end{center}
  \caption{Left: Mass distribution of compact objects in X-ray binaries. 
  Arrows indicate lower limits to BH masses. 
Right: observed mass distribution of compact objects in X-ray binaries 
(shaded histogram), compared to the theoretical distribution computed in 
Fryer \& Kalogera (2001) for the ``Case C + Winds" scenario (dotted line). 
Mass-loss rates by \cite{woos95} were used in the computations. The 
model distribution has been re-scaled for clarity.}
\label{fig:fig4}
\end{figure}

The main property of the BH population is the mass distribution. However,  
in order to get accurate BH masses, in addition to the mass function one needs 
to determine the binary inclination and the mass of the companion star 
or a related quantity such as the mass ratio. 
This information can be extracted from two experiments: (i) resolving the 
rotational broadening $V_{\rm rot} \sin~ i$ of the companion's absorption 
lines, which is correlated with the binary mass ratio  and (ii) fitting 
synthetic models to ellipsoidal lightcurves from which the inclination angle 
can be determined. This is the classic method to derive masses which has
been reviewed in several papers e.g. \cite{casa01}. 

Following this prescription, reliable BH masses have been determined in 15 
binaries. These are listed in the last column of Table 1 and 
displayed in Figure 4, relative to NS masses 
compiled by \cite{stairs04} and \cite{latt04}. 
The figure nicely shows a mass segregation between the 2 populations of 
collapsed objects, with NS clustering around 1.4 M$_{\odot}$ and 
BH masses scattering between 4 and 14 M$_{\odot}$. 
Many more BH masses and errorbars $\le$10\% are needed before fundamental 
questions can be addressed such as 
(i) do BHs masses cluster at a particular value?
(ii) what are the edges of the distribution, i.e the minimum and 
maximum masses for BHs to be  formed in X-ray binaries? 
(iii) is there a continuum distribution of masses between NS and BHs? 
All these questions are intimately related to models of SN and close binary 
evolution.   

Figure 4 also shows the histogram of the current distribution of compact
remnant masses, 
relative to a model distribution computed by \cite{fryer01} which includes 
mass-loss through winds and binarity effects. The model shows a mass cut at 12 
M$_{\odot}$ but this seems to be challenged by the compact objects in V404 Cyg 
and GRS 1915+105, with 
12 $\pm$ 2 M$_{\odot}$ and 14 $\pm$ 4 M$_{\odot}$ respectively. 
However, the conflict may be spurious since recent estimates of mass-loss rates 
in Wolf-Rayet stars suggests that previous determinations were biased too low
(\cite{nugis00}). Also, the model predicts a continuum distribution between NS 
and BH whereas observations seem to show a paucity of objects at 
$\sim$3-4 M$_{\odot}$. This may be caused by selection effects since low-mass 
BHs are expected to be persistent. 
Recently, compact object masses in  the range 2-4 M$_{\odot}$ have been 
reported in 4U1700-37 (\cite{clark02}), V395 Car (\cite{sha04}) and LS 5039 
(\cite{casa05}). 
Although they are assumed to contain NS, none of them has shown X-ray bursts 
nor pulses. Therefore, they may well be members of the 
missing low-mass BH population. We are clearly dominated by low number 
statistics and more observations are required. 

\section{Conclusions}

The best observational evidence for the existence of stellar-mass BHs is 
provided by dynamical studies of X-ray binaries
The first solid {\it candidates} were the classic X-ray binaries 
Cyg X-1 and, in particular, A0620-00. BHs became {\it confirmed} with the 
discovery of mass functions in excess of 5-6 M$_{\odot}$, the first case being 
V404 Cyg. Their global X-ray properties (such as the lack of pulsations/bursts, 
weak quiescent Lx, etc.) also support the presence of an event horizon in these objects.

XRTs are the best hunting ground for new stellar-mass BHs with 17 cases 
currently known and estimated masses between 4-14 M$_{\odot}$. These are  
are only the tip of the iceberg of an estimated dormant population of 
$\sim10^3$ BH binaries and $\sim10^9$ stellar-mass BHs in the Galaxy. 
Clearly many more discoveries and better statistics are essential to derive 
useful constraints on BH formation models. 

New strategies, aimed at  
unveiling more quiescent BH transients, and novel techniques, such as the 
detection of the irradiated donor using high-excitation reprocessed lines, 
need to be exploited.   
\bigskip

\begin{acknowledgments}
I acknowledge useful comments from Phil Charles.  
I'm also grateful for support from the Spanish MCYT grant AYA2002-0036.
\end{acknowledgments}

\newcommand{\myit}[1]{\rm{#1}}

\bigskip

\discuss{Ramesh Narayan}{When you make use of the ellipsoidal model
to determine the binary inclination, how do you avoid contamination of
the stellar light by the disc light?}

\discuss{Jorge Casares}{The example I showed is a ``textbook'' lightcurve
of J1655--40 where the hot F6~III companion star totally dominates the
optical light. However, for the great majority of black hole binaries 
(which contain much cooler companions in shorter orbital periods) the
contamination by the accretion disc light certainly starts to show up
in the form of flaring activity and superhumps. It is commonly assumed 
that the active contribution of the accretion disc light decreases at
longer wavelengths and, hence, IR lightcurves are used to estimate the
orbital inclination.}

\discuss{Tsevi Mazeh}{For HMXB, is it a necessary assumption that the
star fills its Roche lobe? If not, the ellipsoidal effect is difficult
to model.}

\discuss{Jorge Casares}{Roche-lobe filling is not a general assumption 
of the models because HMXBs are normally wind-fed systems. Normally,
the ellipsoidal fits are applied with an extra free parameter which is the
star filling-factor.}

\discuss{Xiaopei Pan}{Physical parameters of
black holes have huge uncertainties. The Space Interferometry Mission
(SIM) can provide high-accuracy determination of inclination, mass,
distance, etc. The precisions of black-hole masses can be directly
determined to 5\%. The inclination precision can reach 2\%, and the
distance of black holes can be measured to better that 2\% by the SIM
project. We hope in more collaborations on black hole issues.}

\discuss{Jorge Casares}{The error budget in black-hole mass determination is
clearly dominated by uncertainties in the inclination angle and indeed
the improvement that the SIM project may provide will be extremely
useful.}

\discuss{Andrew King}{Comment: I would like to reinforce your point about
the black holes we detect being the top of an iceberg. Long-period LMXBs
have wide separations, big discs, and are all transient. However, the
interval between outbursts is so long that we will never detect them at
all. We can detect the endpoints of these LMXBs when they contain
neutron stars as very wide binaries with millisecond pulsars, but not
when they contain black holes.}

\end{document}